\documentclass[aps,prl,amsfonts,amssymb,notitlepage,tightenlines,twocolumn,showpacs]{revtex4-1}
\usepackage[intlimits]{amsmath}
\usepackage{graphicx}

\usepackage{braket}
\usepackage{color}

\begin{document}

\title{Room-temperature superparamagnetism due to giant magnetic anisotropy in Mo$_{S}$ defected single-layer MoS$_{2}$}

\author{M. A. Khan $^{(1,3)}$}
\author{Michael N. Leuenberger $^{(1,2)}$}
\email[e-mail: ]{michael.leuenberger@ucf.edu}
\affiliation{$^{(1)}$ NanoScience Technology Center, Department of Physics, University
of Central Florida, Orlando, FL 32826, USA. \\
$^{(2)}$ College of Optics and Photonics, University of Central Florida, Orlando, FL 32826, USA.\\
$^{(3)}$ Federal Urdu University of Arts, Science and Technology, Islamabad, Pakistan.}

\begin{abstract} 
Room-temperature superparamagnetism due to a large magnetic anisotropy energy (MAE) of a single atom magnet  has always been a prerequisite for nanoscale magnetic devices. Realization of two dimensional (2D) materials such as single-layer (SL) MoS$_{2}$,  has provided new platforms for exploring magnetic effects, which is important for both fundamental research and for industrial applications. Here, we use density functional theory (DFT) to show that the antisite defect (Mo$_{S}$) in SL MoS$_{2}$ is magnetic in nature with a magnetic moment of $\mu$ of $\sim$ 2$\mu_{B}$ and, remarkably, exhibits an exceptionally large atomic scale MAE$=\varepsilon_{\parallel}-\varepsilon_{\perp}$ of $\sim$500 meV. Our calculations reveal that this giant anisotropy is the joint effect of strong crystal field and significant spin-orbit coupling (SOC). In addition, the magnetic moment $\mu$  can be tuned between 1$\mu_{B}$ and 3$\mu_{B}$ by varying the Fermi energy $\varepsilon_{F}$, which can be achieved either by changing the gate voltage or by chemical doping. We also show that MAE can be raised to $\sim$1 eV with n-type doping of the MoS$_{2}$:Mo$_{S}$ sample. Our systematic investigations deepen our understanding of spin-related phenomena in SL MoS$_{2}$ and could provide a route to nanoscale spintronic devices.
\end{abstract}

\maketitle
\begin{figure*}[tp]
           \begin{center}
                      \includegraphics[width=7in]{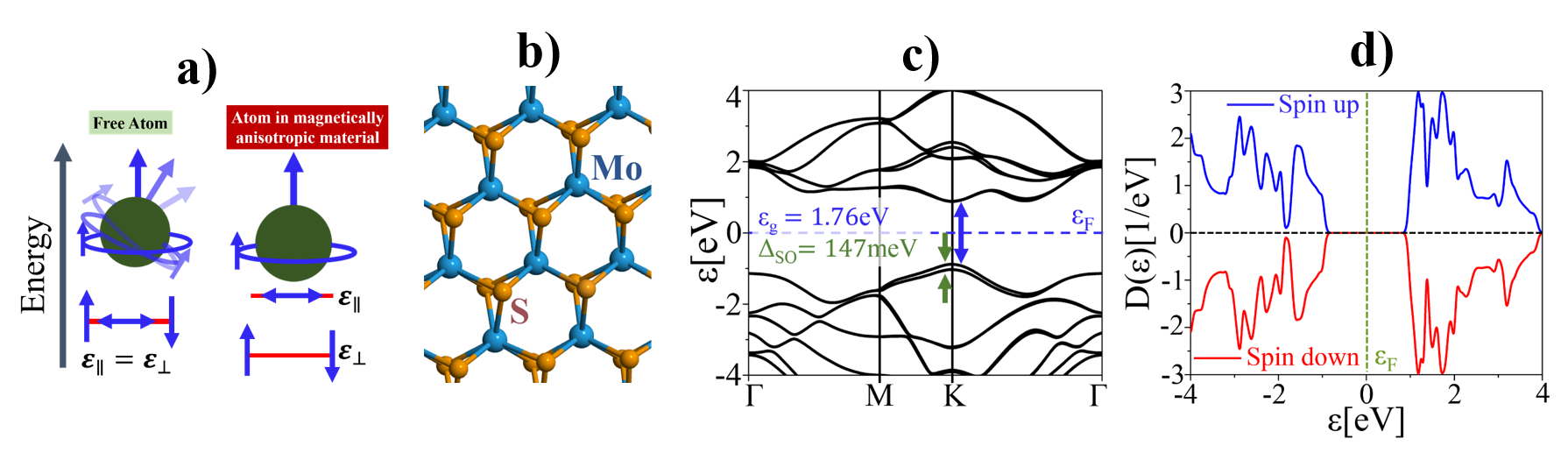}
           \end{center}
            \caption{a) Schematic diagram explaining perpendicular MAE, $=\varepsilon_{\parallel}-\varepsilon_{\perp}$. For an indiviual atom (left) magnetization vector has the same energy for in-plane and out-of-plane directions of magnetization vector i.e. MAE$=0$. While for magnetically anisotropic material (right) energy is required for switching the magnetization vector from out-of-plane to in-plane direction, i.e. MAE$\neq 0$. b) Structure of pristine MoS$_{2}$ with a lattice constant of 3.161\AA. c) Band structure of pristine SL MoS$_{2}$, showing direct band gap of 1.76 eV at K point with SOC of 147 meV . d) Spin-polarized density of states of pristine SL MoS$_{2}$.}
             \label{fig:MAE}
\end{figure*}
\begin{figure}[b]
	\begin{center}
		\includegraphics[width=3.5in]{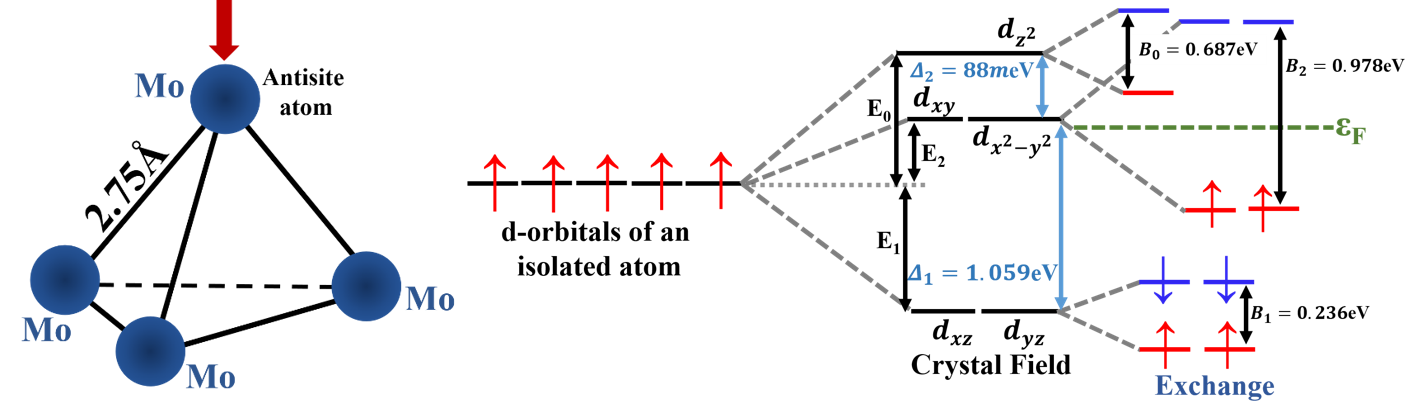}
	\end{center}
	\caption{Trigonal symmetry (left) seen by an antisite atom. Schematic representation of d level splittings of Mo$_{S}$ atom (right) due to the crystal field with C$_{3h}$ point symmetry and exchange interaction.}
	\label{fig:Trigona_Schematic}
\end{figure}

\paragraph{Introduction.}\noindent Single atom magnets adsorbed on the surface of nonmagnetic semiconductors has attracted a great deal of attention over the past few years, as they are potential candidates for the realization of ultimate limit of bit miniaturization for information storgae and processing \cite{Single_Atoms_adsorbed_1,Single_Atoms_adsorbed_2,Single_Atoms_adsorbed_3,Single_Atoms_adsorbed_4,ML}. Superparamagnetsim, usually dominating the magnetic behavior of a single atom magnet, has its origin in the magnetic anisotropy energy (MAE), which measures the energy barrier for flipping the spin moment between two degenerate magnetic states with minimum energy. One of the key factors that limits the performance of nanomagnetic devices is thermal fluctuations of magnetization that eventually randomize the direction of the magnetic state. This loss of information can be prevent by either lowering the operating temperature or by increasing the MAE. It has been shown experimentally that Ho atoms on the surface of MgO exhibit a magnetic remanence up to a temperature of 30 K, corresponding to 2.5 meV, and a relaxation time of 1500 s at 10 K \cite{Ho_MgO}. In Ref.~ \cite{SA_memory} it has recently been demonstrated experimentally that it is possible to read and write a single bit of information using the magnetic state of individual Ho atoms adsorbed on MgO. Remarkably, Ho atoms retain their magnetic information over many hours at 1.2 K. It is therefore highly desireable to engineer materials with as large as possible MAEs to produce stable magnetization above room tmperatures. It has also been shown that adatoms of Co, Ru and Os on the surface of MgO shows MAE $\sim$ 100 meV \cite{Exp_MAE_1,Th_MAE_2}. All these efforts require deposition of transition metal atoms on the surface of non-magnetic semiconductors, such as MgO. Here, by using density functional theory (DFT) we show that an exceptionally large MAE $\sim$ 500 meV can be observed in SL MoS$_{2}$ in the presence of an antisite (Mo$_S$, Mo replacing S) defect.\newline 
The concept of MAE, which is a preferential spatial orientation of magnetization, is not relevant in an individual isolated atom \cite{Khajetoorians976,Yosida}, i.e., magnetic moments freely rotate in any direction without energy cost (Fig.~\ref{fig:MAE}(a)). However, for an adsorbed or impurity atom, crystal field effects localize the electrons to the directional bonds and effectively quench the orbital motion. SOC tries to restore partially this quenching of orbital angular momentum and ultimately leads to magnetic anisotropy.\newline
Two dimensional (2D) materials are generally catagorized as 2D allotrophes of various elements or compounds, in which electron motion is confined to a plane such as graphene, phosphorene and SL MoS$_{2}$. Apart from their fascinating electronic and optical properties, 2D materials are very attractive for spintronic applications \cite{2D_Spin_1,2D_Spin_2,2D_Spin_3,2D_Spin_4,2D_SP_5,2D_SP_6,2D_SP_7}. From a technological perspective 2D materials have advantages that can be employed in magnetic and spintronic devices. First, 2D materials provide an excellent control over carrier concentration through gate voltage. Secondly, it has been shown that carrier density in 2D materials is relatively stable against thermal fluctuations \cite{Robust_th_Fluc}.\newline
SL MoS$_{2}$ is a direct band gap semiconductor with considerable SOC ($\sim$150meV) that originates from the d-orbitals of heavy Mo atoms and due to the lack of inversion symmetry. High quantum efficiency \cite{Quantum_efficiency_1,Quantum_efficiency_2}, acceptable value for the electron mobility \cite{electron_mobility_1,electron_mobility_2} and low power dessipation \cite{low_power_dissipation_1,low_power_dissipation_2}, makes MoS$_{2}$ a candidate material for future electronic devices.  Despite its success as a fascinating SL semiconductor, magnetism in MoS$_{2}$ has remained almost unexplored.\newline
Different fabrication techniques, such as physical vapor deposition (PVD), chemical vapor deposition (CVD) and mechanical exfoliation, have been used to produce wafer scale chunks of MoS$_{2}$. In Ref.~\onlinecite{Antisite_defects} it has been shown that the abundance of defects present in MoS$_{2}$ depends on the fabrication method. In particular, the large abundance of Mo$_{S}$ defects has been observed in PVD-grown MoS$_{2}$. Although Mo$_{S}$ defects have been explored both experimentally and theoretically in terms of the electronic structure \cite{Antisite_defects,Antisite_defects_1}, a comprehensive investigation regarding magnetic behavior is still missing.\newline
Here, we present a comprehensive study based on DFT calculations to show that Mo$_{S}$ defects are magnetic in nature. In particular, we show that a sizeable localized magnetic moment ($\mu\sim2\mu_{B}$) is associated with an Mo$_{S}$ defect in MoS$_{2}$. In addition, $\mu$ can be tuned by changing the carrier concentration (or Fermi level), which can be achieved either by gate voltage or by doping. Remarkably,we show that antisite defects in MoS$_{2}$ possess exceptionally large MAE. Our calculations reveal that MAE originates from the combined effect of strong crystal field and SOC in MoS$_{2}$. This large value of MAE will be of major interest for applications in which the axial states representing an information bit need to be protected from thermal fluctuations at room temperature.
\paragraph{Numerical results.} 
All numerical calculations have been carried out using DFT and with the use of Perdew-Burke-Ernzerh (PBE) generalized gradient (GGA) parametrization for exchange-correlation functional. Both spin-polarized and relativistic SOC calculations are performed. The sampling of the Brillouin zone was done for a supercell with the equivalent of a 35$\times$35$\times$1 Monkhorst–Pack k-point grid for the MoS$_{2}$  primitive unit cell with a cutoff energy of 300 Ry. For all calculations, structures are first geometrically optimized with a force tolerance of 0.005 eV/\AA. The calculations are implemented within Atomistic Toolkit 2016.1 \cite{QW_1}. We first obtain the results for the band structure and the density of states for pristine MoS$_{2}$, as shown in Fig.~\ref{fig:MAE}(c) and (d), respectively. Band gaps, SOC and lattice constant (3.161\AA) values for MoS$_{2}$  are in good agreement with previously reported values \cite{Coupled_valley_spin,Three_Band_TBM}. The curves of DOS  (Fig.~\ref{fig:MAE}(d)) for spin-up and spin- down electrons are totally symmetric and the Fermi level is located in the band gap region, suggesting that  pristine MoS$_{2}$ is a nonmagnetic semicondutor. \newline
\begin{figure*}[tb]
	\begin{center}
		\includegraphics[width=7in]{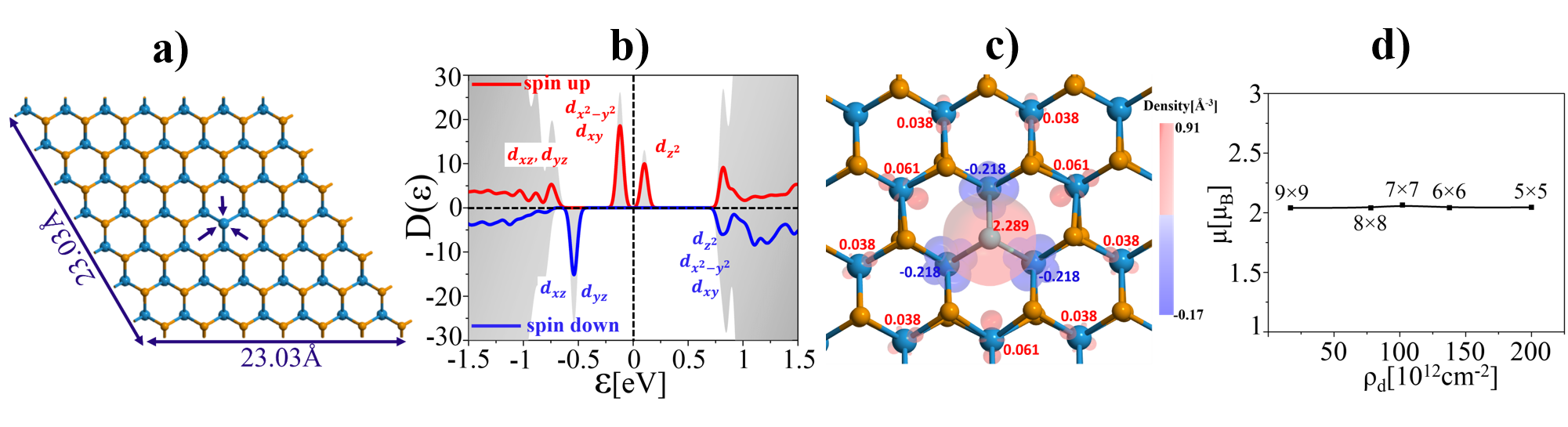}
	\end{center}
	\caption{(a) 7$\times$7$\times$1 supercell with Mo$_{S}$ defect used for calculations (b) Spin-polarized density of states for an Mo$_{S}$ in a 7$\times$7$\times$1 super cell of MoS$_{2}$. Red(blue) is for spin-up(down) projected DOS for Mo$_{S}$ atom and gray color shows the total DOS. (c) Electron difference density $\rho_{\uparrow}-\rho_{\downarrow}$, with MP values (numbers) (d) Magnetic moment $\mu$ vs Mo$_{S}$ defect density $\rho_{d}$.}
	\label{fig:SP_MoS2}
\end{figure*}
For the Mo$_{S}$ defect calculations,  we consider a 7$\times$7$\times$1 supercell with an edge length of 23.03 \AA [see  Fig.~\ref{fig:SP_MoS2}(a)]. The point group of MoS$_{2}$ with Mo$_{S}$ defect is C$_{3v}$ and it remains preserved after geometrical optimization. The magnetic energy gain $\Delta\varepsilon=0.55$eV, which is the difference in ground states energy $\Delta\varepsilon=\varepsilon_{NSP}-\varepsilon_{SP}$ between the non-spin-polarized (NSP) and spin-polarized (SP) calculations, indicates that the paramagnetic phase is stable well above room temperature. To visualize the magnetic properties resulting from the Mo$_{S}$ defect we plot the SP DOS (Fig.~\ref{fig:SP_MoS2}(b)) corresponding to the configuration shown in Fig.~\ref{fig:SP_MoS2}(a). Fig~\ref{fig:SP_MoS2} (b) shows a significant change in the spin-up and spin-down total DOS (grey) as compared with pristine MoS$_{2}$ (Fig~\ref{fig:MAE}(d)). To understand the origin of this change, we plot the SP projected DOS at the Mo$_{S}$ atom (Fig.~\ref{fig:SP_MoS2}(b)), which shows that SP is induced mainly due to the Mo$_{S}$ defect. For further illustration we show the results for the SP isosurface and the Mulliken Population (MP)\cite{Mulliken_population} analysis [Fig.~\ref{fig:SP_MoS2} (c)]. Fig.~\ref{fig:SP_MoS2}(c) shows that magnetic moment resides mainly on the Mo$_{S}$ atom, decays sharply, and becomes negligibly small beyond a few lattice constants. Magnetic moment associated with Mo$_{S}$ defect in MoS$_{2}$ is found to be 2.04$\mu_{B}$. When an impurity atom is put into a crystal environment, crystal field effects break the orbital degeneracies of the impurity atom. An  Mo$_{S}$ defect in MoS$_{2}$ sees a trigonal crystal field [Fig.~\ref{fig:Trigona_Schematic} (a)], for which the energy level diagram is shown in Fig.~\ref{fig:Trigona_Schematic}(b). The crystal field splitting associated with the C$_{3h}$ symmetry seen by the Mo$_{S}$ defect lifts the degeneracy of the d-orbitals of the Mo$_{S}$ defect and splits them into three multiplets $e^\prime$ (d$_{x^2-y^2}$ and d$_{xy}$ orbitals), $e^{\prime\prime}$ (d$_{xz}$ and d$_{yz}$) and $a^\prime_{1}$ (d$_{z^2}$). The exchange interaction then leads to further splitting for the states with the opposite spins. The total spin should be governed by the unpaired spin counts according to Hund's rules. In solids the Fermi level plays a decisive role in populating or depopulating certain atomic levels. Hund's rules together with the position of the Fermi level predict a magnetic moment of 2$\mu_{B}$, which is an excellent agreement with the values obtained by means of our numerical results (Fig.~\ref{fig:SP_MoS2}(c)). \newline
The 2D nature of MoS$_2$ provides the possibility of gating and thereby controlling both the electrical and magnetic properties by tuning the carrier density. The Fermi level of 2D materials can be shifted by changing the gate voltage or by doping. It has been shown that \cite{donor_acceptor_levels,Chloride_doping} substitutional doping with the S atom replaced by a Cl (P) atom leads to n(p)- type doping in SL MoS$_{2}$. To develop a connection between magnetic moment and carrier density, we consider a 7$\times$7$\times$1 supercell containing an Mo$_{S}$ defect and a substitutional Cl (P) atom as an n(p)- type dopant. We find that the magnetic moment due to an Mo$_{S}$ defect can be increased to 3$\mu_{b}$ or decreased to $\sim$1$\mu_{B}$) by raising or lowering the Fermi level, respectively. The tunablility of the magnetic moment by electrical means is highly desireable from fundamental and technological perspectives, especially in view of recent developments in magnetoelectronics and spintronics \cite{magnetic_moment_controle,TBM_MM,Prinz1660, Nat_electrical_controle}.\newline
In Fig.~\ref{fig:SP_MoS2} (d) we plot the magnetic moment vs various defect densities. It can be seen that the magnetic moment does not change for different concentrations of Mo$_{S}$ defects, which shows that the magnetic moment is localized and does not interact with neighboring defects. Therefore there is no ferromagnetic or antiferromagnetic ordering.\newline
\begin{figure}[b]
	\begin{center}
		\includegraphics[width=3.5in]{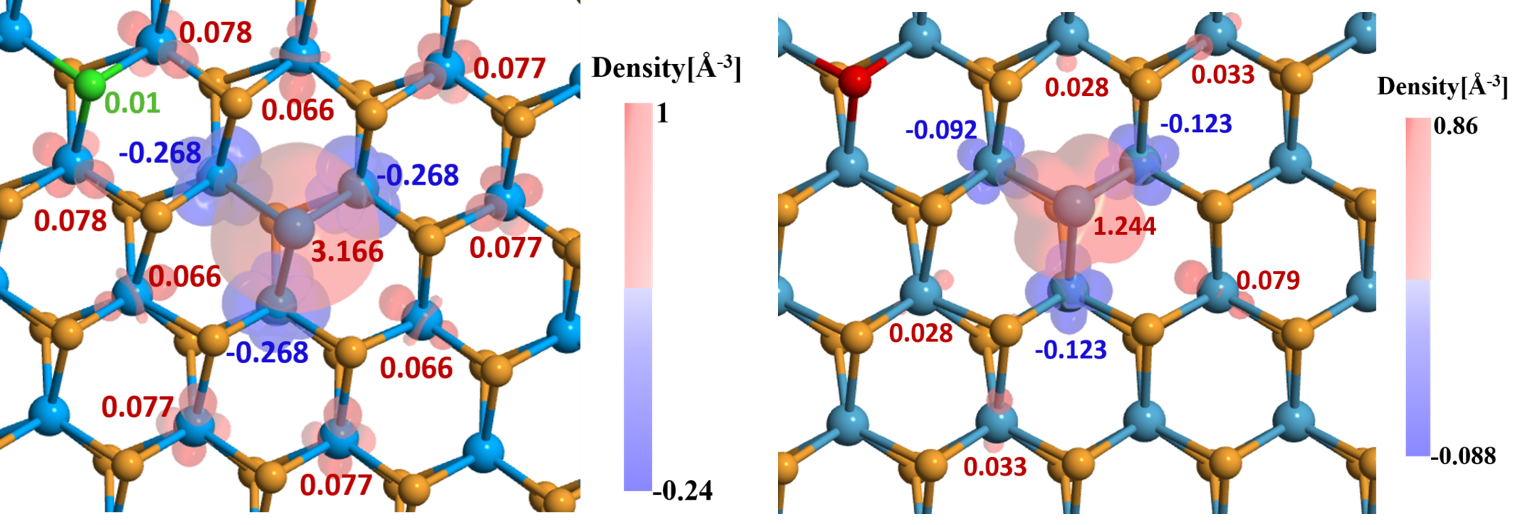}
	\end{center}
	\caption{Electron difference density  $\rho_{\uparrow}-\rho_{\downarrow}$ for n-doped (left) and p-doped (right) with a doping concentration of 32$\times$10$^{12}$cm$^{-2}$. Green(red) circle shows Cl(P) atoms. MP values (numbers) are also shown.}
	\label{fig:spin_density}
\end{figure}
\begin{figure}[t]
           \begin{center}
                      \includegraphics[width=3.5in]{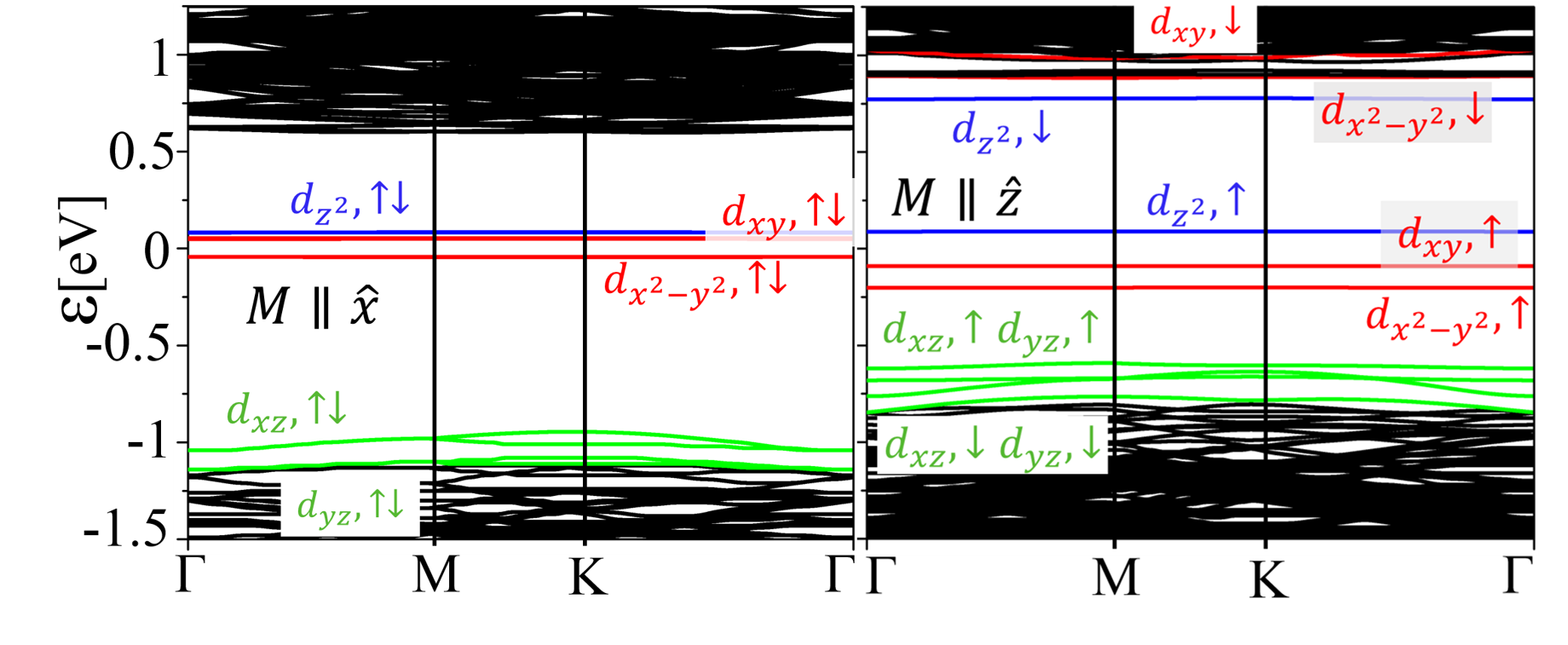}
           \end{center}
\caption{Band structures of 7$\times$7$\times$1 super cell of SL MoS$_{2}$ with Mo$_{S}$ defect including SOC with M$\parallel\hat{x}$(left) and M$\parallel\hat{z}$(right). Black lines show extended electronic states, colored lines show the localized defect states of Mo$_{S}$ atom.}
\label{fig:BS_complete}
\end{figure} 
The MAE value is calculated by employing a two-step process. First, a Kohn-Sham based calculation with collinear electron density and without SOC corrections is performed in order to obtain a self-consistent ground state electronic charge density. In the second step, the obtained charge density is used as an input for the non self-consistent SOC and non-collinear calculations with varying orientation of the magnetic moments. We consider two magnetization directions, i.e. in-plane ($\parallel$) and out-of-plane ($\perp$) to the 2D sheet of MoS$_{2}$. The energy difference $\epsilon_{\parallel}$-$\epsilon_{\perp}$ calculated by using SGGA+SOC calculations shows that the MAE can be as large as 550 meV ($S=1$) per Mo$_{S}$ defect, with highly preferential easy axis pointing out-of-plane. It is well known that higher values of magnetization lead to larger anisotropy. SGGA+SOC calculations for an n-doped MoS$_2$:Mo$_S$ sample (Fig.~\ref{fig:spin_density}(a)) show that the perpendicular MAE can be as large as 980 meV ($S=3/2$). It is important to mention that our calculations show that there is no preferential in-plane orientations of the magnetization, which means that our system is described by an easy axis only. Zero field splitting Hamiltonian for a single atom magnet can be written as
\begin{equation}\label{eq:Hamiltonian_SAM}
\hat{H} = D(\hat{S}_{z}^{2}-\frac{1}{3}S(S+1)),
\end{equation} 
where $D$ is related to the unquenched orbital angular momentum along the local axial direction of the magnetic ion. If $D<0$ axial spin states are preferred over the planar ones, which means that the spin is aligned with respect to the $z$-axis, defining the easy axis. For $S=1$ and $S=3/2$ the corresponding level splittings are $|D|$ and $2|D|$, respectively. This simplified model fits the numerical results for a value of $D=-510$ meV with deviations of $\pm 40$ meV. \newline
To investigate the effects of SOC on the magnetization direction, we plot band stuctures in the presence of  SOC, with in-plane and out-of-plane magnetization directions in Fig.~\ref{fig:BS_complete}. Fig.~\ref{fig:BS_complete}(a) and (b) show that the influence of the SOC is significantly larger for M$\parallel\hat{z}$ than for M$\parallel\hat{x}$. More specifically, Kramers degeneracy, which arises due to time reversal symmetry, is preserved for M$\parallel\hat{x}$, while it is broken for M$\parallel\hat{z}$. This contrast in the band structures can be linked to the MAE of the Mo$_{S}$ atom. It is well known that magnetic ordering such as ferromagnetism or more related (in the context of single ion) superparamagnetism, breaks the time reversal symmetry, which in turn lifts the Kramers degeneracy. Our DFT calculations reveal that for M$\parallel\hat{x}$ Kramer doublets remain degenerate, indicating paramagnetism, while for M$\parallel\hat{z}$ Kramers degeneracy is lifted, which is due to superparamagnetism. The large energy barrier between out-of-plane and in-plane magnetization directions shows that superparamagnetism is more stable than paramagnetism well above room temperatures.
\paragraph{Analytical modeling.} We see that SOC splits the electronic states for different orientations of the magnetization, thereby creating the large anisotropy. To understand up to what extent this can be explained analytically, we present a simple analytical model \cite{analytical_model} that systematically considers all the essential factors contributing to the MAE, i.e. the crystal field effect $\hat{H}^{cry}$, the exchange field effect $\hat{H}^{excch}$, and the SOC $\hat{H}^{SOC}$. The simplified model Hamiltonian can be written as
\begin{equation}\label{eq:Hamiltonian}
\hat{H} = \hat{H}^{cry}+\hat{H}^{exch} + \hat{H}^{SOC}.
\end{equation}
\begin{figure}[b]
	\begin{center}
		\includegraphics[width=3.5in]{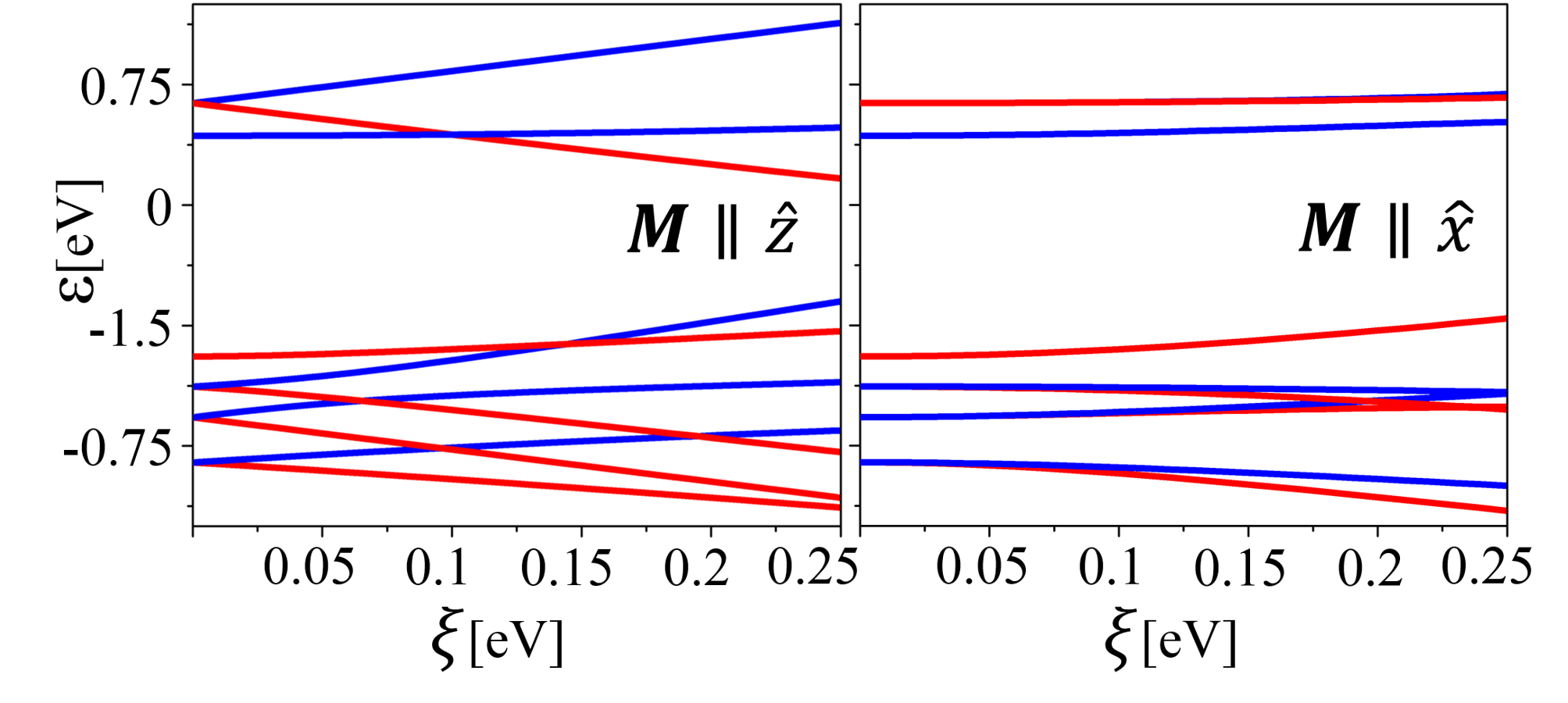}
	\end{center}
	\caption{Dependence of eigen energies of the model Hamiltonian given by Eq.~\ref{eq:Hamiltonian} on the SOC strength $\xi$ for M$\parallel\hat{z}$ (left) and M$\parallel\hat{x}$  (right).}
	\label{fig:analytical_splitting}
\end{figure}
To highlight the essential features, we only consider the d$-$orbitals of the Mo$_{S}$ atom. In crystal field theory the key factor is to find an expression for the field produced by point charges which possess a given symmetry. An Mo$_{S}$ atom sees a trigonal electrostatic environment due to the nearest neighbour (NN) Mo atoms of MoS$_{2}$. The crystal field hamiltonian $\hat{H}^{cry}$ describing the electrostatic field produced by the NN Mo atoms, at Mo$_{S}$ site is $\sim$$Y_{2}^{0}$, where $l=2$ and $m=0$ are orbital and magnetic quantum numbers, respectively. The crystal field Hamiltonian lifts the d$-$ orbital degeneracy of the Mo$_{S}$ atom by forming two doublets $d_{xy}$/$d_{x^2-y^2}$($m=\pm2$), $d_{xz}$/$d_{yz}$($m=\pm1$) and a singlet d$_{z^2}$($m=0$), which is in agreement with our numerical results. Considering the fact that crystal field theory preserves the level splittings with respect to the degenerate d-orbitals of the isolated Mo atoms, i.e. $E_{0}+2E_{1}+2E_{2}=0$, the eigenenergies of the crystal field Hamiltonian can be written in the form of energy differences $\Delta_{1}$ and $\Delta_{2}$, as shown in Fig.~\ref{fig:Trigona_Schematic}.\newline
For the exchange Hamiltonian we consider the spin quantization axis fixed (parallel to $\hat{z}$-axis). For magnetization M$\parallel\hat{z}$
\begin{equation}\label{eq:exchange_Hamiltonian_z}
\begin{split}
\hat{H}^{exch}_{m^{\prime}s^{\prime},ms}=B_{m},\quad m=m^{\prime}, \quad s=s^{\prime}=1/2,\\
\hat{H}^{exch}_{m^{\prime}s^{\prime},ms}=-B_{m},\quad m=m^{\prime},\quad s=s^{\prime}=-1/2,
\end{split}
\end{equation}
where the subscript $m$ shows that the exchange splitting field $B_{m}$ depends on the magnetic quantum numbers (Fig.~\ref{fig:Trigona_Schematic}(b)). For M$\parallel\hat{x}$
\begin{equation}\label{eq:exchange_Hamiltonian_x}
\hat{H}^{exch}_{m^{\prime}s^{\prime},ms}=B_{m},\quad m=m^{\prime},\quad s\neq s^{\prime}.
\end{equation}
The third contribution to the model Hamiltonian comes from the SOC. SOC is considered as the onsite interaction $\hat{H}^{SOC}=\xi L\cdot S$. The effect of the SOC inducing the splittings in energy levels can be obtained by diagonalizing the Hamiltonian (\ref{eq:Hamiltonian}) for M$\parallel\hat{x}$ and M$\parallel\hat{z}$. The values of the various parameters $\Delta_{1}$, $\Delta_{2}$, B$_{0}$, B$_{1}$ and B$_{2}$ are extracted from numerical calculations (Fig.~\ref{fig:Trigona_Schematic}), and the corresponding results are presented in Fig.~\ref{fig:analytical_splitting}. A pertinent feature of Fig.~\ref{fig:analytical_splitting} is that the effect of $\xi$ is much weaker for in-plane magnetization M$\parallel\hat{x}$ than for the out-of-plane magnetization M$\parallel\hat{z}$, which is in agreement with our numerical results. Specifically, our simple analytical model shows that eigenenergies remain 2-fold degenerate (Kramers doublet) for sufficiently high values of SOC parameter $\xi$ for M$\parallel\hat{x}$, whereas degeneracy is completely lifted for M$\parallel\hat{z}$. The simple analytical model qualitatively explains the time reversal symmetry breaking for M$\parallel\hat{z}$, identifying the superparamagnetic state and also indicating that the MAE can be understood as the interplay between the crystal field, the exchange field, and SOC.
\paragraph{Conclusion.} We have demonstrated that an Mo$_{S}$ defect in MoS$_{2}$ carries a magnetic moment of $\mu_{B}$, 2$\mu_{B}$, and 3$\mu_{B}$, which can be tuned by changing the position of Fermi level electrostatically. Remarkably, an Mo$_{S}$ defect in MoS$_{2}$ exhibits an exceptionally large MAE of 550 meV with out-of-plane easy axis. Our calculations reveal that this very large anisotropy is the combined effect of strong crystal field and SOC. We show that the MAE can be tuned up to $\sim$1 eV with n-type doping, which allows for room-temperature operation of future magnetic memory devices based on single atomic defects. 
     
\begin{acknowledgments}
M.L. acknowledges support provided by NSF grant CCF-1514089.  
\end{acknowledgments}

\section{Appendix}
\begin{figure}[b]
	\begin{center}
		\includegraphics[width=3.5in]{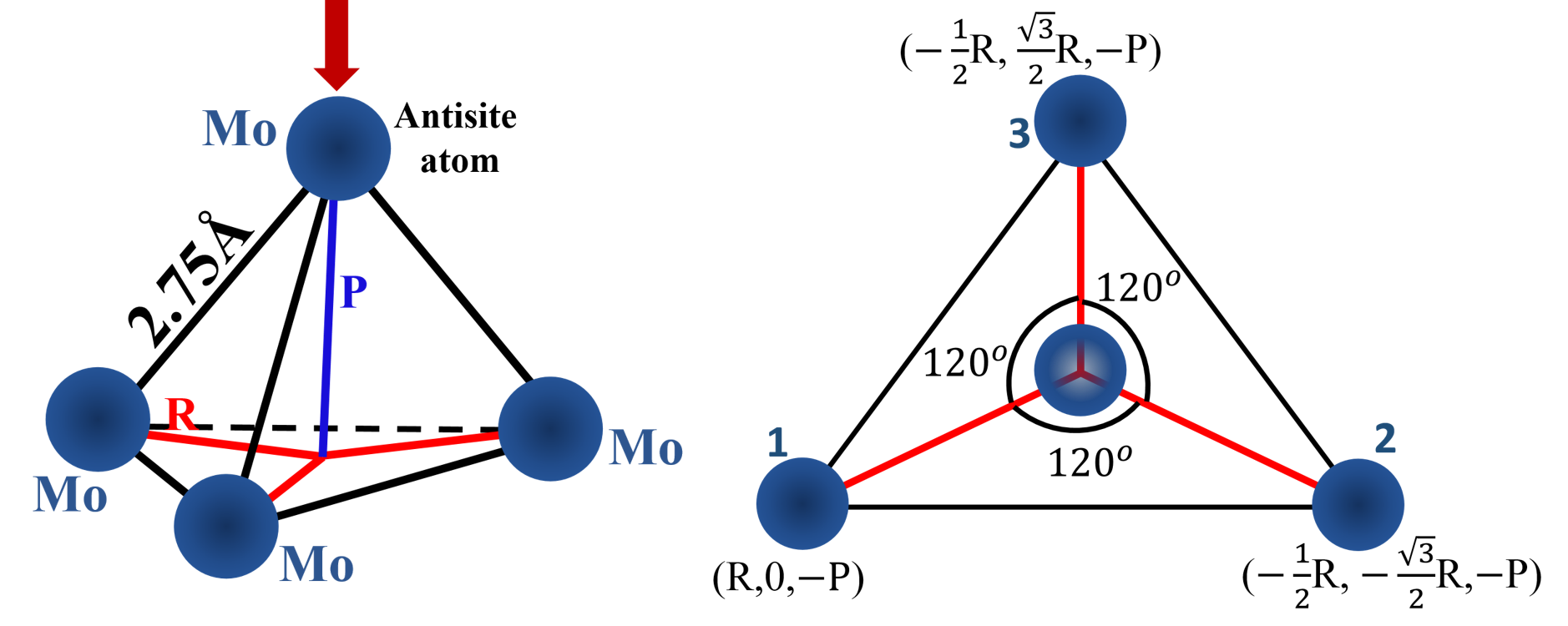}
	\end{center}
	\caption{Trigonal symmetry seen by the Mo$_S$ atom. The origin is set at the Mo$_{S}$ atom. One of the Mo atom is set at x-axis and the coordinates of the 2 and 3 atoms are obtained through rotation of coordintes.}
	\label{fig:appendix_pic}
\end{figure}

We derive here the crystal field Hamiltonian for trigonal symmetry (Fig.~\ref{fig:appendix_pic}). The contribution of the surroundings point charges (Mo atoms, Fig.~\ref{fig:appendix_pic}) to the electron potential energy at Mo$_{S}$ site can be expressed as
\begin{equation}\label{eq:point_charge}
V_{CF}=\displaystyle\sum_{i=1}^{3} \frac{Ze^2}{|\vec{r}-\vec{R}_{i}|}
\end{equation}
where $\vec{r}$ is the electron corrdinate and $\vec{R}_{i}$ are the position vectors of the neighboring point charges. With the help of Mathematica \cite{Mathematica} we can write down the expression for the crystal field Hamiltonian  
\begin{equation}\label{eq:point_charge}
\begin{split}
V_{CF}=C_{0}+C_{1}\rho Y_{1}^0+C_{2}\rho^2Y_{2}^0+\rho^3[C_{3}Y_{3}^0\\
+C_{3}^{\prime}(Y_{3}^{-3}+Y_{3}^{3})]+\rho^4[C_{4}Y_{4}^0+C_{4}^{\prime}((Y_{4}^{-3}+Y_{4}^{3}))]....,
\end{split}
\end{equation}
where $\rho=r/\sqrt{R^2+P^2}$ and $Y_l^m$ are the spherical harmonics with orbital angular momentum quantum numbers $l$ and $m$. The expansion coefficients $C_j$, $j=0,1,2,...$ can be adjusted to fit the DFT results. Here we use $d$-orbitals of the Mo$_{S}$ atom, i.e. $d_{x^2-y^2}=(Y_{2}^{-2}+Y_{2}^{2})/\sqrt{2}, d_{xy}=i(Y_{2}^{-2}-Y_{2}^{2})/\sqrt{2},d_{z^2}=Y_{2}^0,d_{xz}=(Y_{2}^{-1}+Y_{2}^{1})/\sqrt{2}$ and $d_{yz}=i(Y_{2}^{-1}-Y_{2}^{1})/\sqrt{2}$. Spherical harmonics with odd magnetic quantum numbers do not contribute, thus $V_{CF}\sim \rho^{2}Y_{2}^{0}$ in lowest order. The matrix elements of the $V_{CF}$ between different $d$-orbitals may be written as
\begin{equation}\label{eq:matrix_elements}
\hat{H}_{mm^{\prime}}^{cry}\sim\int\psi_{nl}^{*}(r)\rho^{2}\psi_{nl}(r)r^{2}dr\iint d_{m}(\theta,\phi)Y_{2}^{0}d_{m^\prime}(\theta,\phi)d\theta d\phi
\end{equation}
where $\psi_{nl}(r)$ is the radial function for Mo$_{S}$ atom ($n=4$, $l=2$) and subscripts $m$ and $m^{\prime}$ stand for different $d$-orbitals of the Mo$_{S}$ atom. In this work we are able to omit the radial parts by fitting the appearing integrals, this spatial distribution may be omitted, wich allows to simplify the treatment with any loss of accuracy. The diagonal matrix elements are given by
\begin{equation}\label{d_orbital}
\begin{split}
\bra{d_{z^2}}Y_{2}^{0}\ket{d_{z^2}}=145\sqrt{5\pi}/512=E_0, \\ 
\bra{d_{x^2-y^2}}Y_{2}^{0}\ket{d_{x^2-y^2}}=-45\sqrt{5\pi}/1024=E_2, \\
\bra{d_{xy}}Y_{2}^{0}\ket{d_{xy}}=-45\sqrt{5\pi}/1024=E_2, \\
\bra{d_{xz}}Y_{2}^{0}\ket{d_{xz}}=-15\sqrt{5\pi}/256=E_1, \\
\bra{d_{yz}}Y_{2}^{0}\ket{d_{yz}}=-15\sqrt{5\pi}/256=E_1. \\
\end{split}
\end{equation}
It should be noted that all the of diagonal terms are zero with in the lowest  approximation ($V_{CF}\sim Y_{2}^0$). Eq.~(\ref{d_orbital}) correctly reproduces the numerical results, i.e. two doublets $d_{x^2-y^2}/d_{xy}$, $d_{xz}/d_{yz}$ and a singlet $d_{z^2}$ with the correct energy sequence $E_{0}>E_{2}>E_{1}$.\newline
Considering the fact that crystal field theory preserves the level splittings with respect to the degenerate d-orbitals of the isolated Mo atoms, i.e. $E_{0}+2E_{1}+2E_{2}=0$, the eigenenergies of the crystal field Hamiltonian can be written in the form of energy differences $\Delta_{1}=E_2-E_1$ and $\Delta_{2}=E_0-E_2$ with $E_0>0$, $E_1<0$, $E_2>0$, as shown in Fig.~\ref{fig:Trigona_Schematic}. The crystal field Hamiltonian may be written as\newline
\begin{widetext}
\begin{equation}
   \hat{H}^{cry} = \left(
		\begin{array}{ccccc}
		\frac{1}{5}(2\Delta_{1}-\Delta_{2})&0&0&0&0\\
		0&\frac{1}{5}(2\Delta_{1}-\Delta_{2})&0&0&0\\
                     0&0&\frac{2}{5}(\Delta_{1}+2\Delta_{2})&0&0\\
                     0&0&0&-\frac{1}{5}(3\Delta_{1}+\Delta_{2})&0\\
                     0&0&0&0&-\frac{1}{5}(3\Delta_{1}+\Delta_{2})\\  
           \end{array}
	 \right),
\end{equation}
\end{widetext}
SOC is considered as the onsite interaction $\hat{H}^{SOC}=\xi L\cdot S$. Using the d-orbital bases $\ket{d_{x^2-y^2},\uparrow}, \ket{d_{xy},\uparrow},\ket{d_{z^2},\uparrow},\ket{d_{xz},\uparrow},\ket{d_{yz},\uparrow}$ and $\ket{d_{x^2-y^2},\downarrow}, \ket{d_{xy},\downarrow},\ket{d_{z^2},\downarrow},\ket{d_{xz},\downarrow},\ket{d_{yz},\downarrow}$, we get the SOC contribution to the Hamiltonians $\hat{H}^{SOC}(\vec{M}\parallel \hat{z})$ and $\hat{H}^{SOC}(\vec{M}\parallel \hat{x})$ as
\begin{widetext}
\begin{equation}
   \hat{H}^{SOC}(\vec{M}\parallel \hat{z}) =
 \left(
		\begin{array}{cccccccccc}
		0 & -2i\xi & 0 & 0 & 0 & 0 & 0 & 0 & \frac{\xi}{2} & i\frac{\xi}{2}  \\
		2i\xi & 0 & 0 & 0 & 0 & 0 & 0 & 0 & - i\frac{\xi}{2} & \frac{\xi}{2}  \\
                     0 & 0 & 0 & 0 & 0 & 0 & 0 & 0 & \frac{\sqrt{3}}{2}\xi & -i\frac{\sqrt{3}}{2}\xi  \\
                     0 & 0 & 0 & 0 & -i\xi & \frac{\xi}{2} & -i\frac{\xi}{2} &\frac{\sqrt{3}}{2}\xi & 0 & 0  \\
                     0 & 0 & 0 & i\xi & 0 & i\frac{\xi}{2} & \frac{\xi}{2} & -i\frac{\sqrt{3}}{2}\xi & 0 & 0  \\
                     0 & 0 & 0 & \frac{\xi}{2} & -i\frac{\xi}{2} & 0 & -2i\xi & 0 & 0 & 0  \\
                     0 & 0 & 0 & i\frac{\xi}{2} & \frac{\xi}{2} & 2i\xi & 0 & 0 & 0 & 0  \\
                     0 & 0 & 0 & \frac{\sqrt{3}}{2}\xi & i\frac{\sqrt{3}}{2}\xi & 0 & 0 & 0 & 0 & 0   \\
                     \frac{\xi}{2} & i\frac{\xi}{2} & \frac{\sqrt{3}}{2}\xi & 0 & 0 & 0 & 0 & 0 & 0 & -i\xi  \\
                     -i\frac{\xi}{2} & \frac{\xi}{2} & i\frac{\sqrt{3}}{2}\xi & 0 & 0 & 0 & 0 & 0 & i\xi & 0 \\  
           \end{array}
	 \right)
\end{equation}
\end{widetext}
and
\begin{widetext}
\begin{equation}
   \hat{H}^{SOC}(\vec{M}\parallel \hat{x}) = \left(
		\begin{array}{cccccccccc}
		0 & 0 & 0 & \xi & 0 & 0 & 0 & 0 & 0 & 0  \\
		0 & 0 & 0 & 0 & \xi & 0 & 0 & 0 & 0 & 0  \\
                     0 & 0 & 0 & \sqrt{3}\xi & 0 & 0 & 0 & 0 & 0 & 0  \\
                     \xi & 0 & \sqrt{3}\xi & 0 & 0 & 0 & 0 & 0 & 0 & 0  \\
                     0 & \xi & 0 & 0 & 0 & 0 & 0 & 0 & 0 & 0  \\
                     0 & 0 & 0 & 0 & 0 & 0 & 0 & 0 & \xi & 0  \\
                     0 & 0 & 0 & 0 & 0 & 0 & 0 & 0 & 0 & \xi  \\
                     0 & 0 & 0 & 0 & 0 & 0 & 0 & 0 & \sqrt{3}\xi & 0  \\
                     0 & 0 & 0 & 0 & 0 & \xi & 0 & \sqrt{3}\xi & 0 & 0  \\
                     0 & 0 & 0 & 0 & 0 & 0 & \xi & 0 & 0 & 0  \\  
           \end{array}
	 \right),
\end{equation}
\end{widetext}
respectively.

\end{document}